\documentclass[%
reprint,
 amsmath,amssymb,
 aps,
]{revtex4-2}

\usepackage{float}
\usepackage{slashed}
\usepackage{xcolor}
\usepackage{amsmath}
\usepackage{subcaption}
\usepackage[english]{babel}
\usepackage{lineno, blindtext}
\usepackage{tikz}
\usepackage{adjustbox}
\usepackage{rotating}
\usepackage{rotfloat}
\usepackage{verbatim}
\usepackage[utf8]{inputenc}
\usepackage{graphicx}
\usepackage{dcolumn}
\usepackage{bm}
\usepackage[super]{nth}
\usepackage{lipsum}

\newcommand\Tdiag[4]{%
    \multicolumn{1}{|p{#2}|}{\hskip-\tabcolsep
    \begin{tikzpicture}[%
                baseline={(0,-.25\baselineskip)},
                every node/.style={outer sep=0pt,inner sep=#1}]
    \node[minimum width={#2+1\tabcolsep-\pgflinewidth},
        minimum height=2\baselineskip-\pgflinewidth+\extrarowheight,
        use as bounding box] (box) {};
    \draw[line cap=round] (box.north west) -- (box.south east);
    \node[anchor=south west,text width=.75*#2,align=left] at (box.south west) {#3};
    \node[anchor=north east,text width=.75*#2,align=right] at (box.north east) {#4};
\end{tikzpicture}\hskip-\tabcolsep}}

\usepackage{hyperref}
\usepackage{xcolor}
\hypersetup{
  colorlinks   = true, 
  urlcolor     = red, 
  linkcolor    = blue, 
  citecolor   = blue 
}

\usepackage{threeparttable}
\begin{document}

\preprint{APS/123-QED}

\title{Search for the production of dark fermion candidates in association with heavy neutral
gauge boson decaying to dimuon in proton-proton collisions at $\sqrt{s} = 8$ TeV using the CMS open data}



\author{Y. Mahmoud}
 \altaffiliation[yehia.abdelaziz@bue.edu.eg]{}
\affiliation{%
Centre for theoretical physics, The British University in Egypt,  
Physics Department, Faculty of Science, Cairo University. 
}%

\author{H. Abdallah and M. T. Hussein}
\affiliation{%
 Physics Department, Faculty of Science, Cairo University. 
}%

\author{S. Elgammal}%
\affiliation{%
Centre for theoretical physics, The British University in Egypt.  
}

\date{\today}

\begin{abstract}
In this paper, we have done a search for dark matter using a part of the data recorded by the CMS experiment during run-I of the LHC in 2012 with 8 TeV center of mass energy and integrated luminosity of 11.6 fb$^{-1}$. 
This data has been gathered from the CMS open data. 
The dark matter, in the framework of simplified model (mono-Z$^{\prime}$), 
can be produced from proton-proton collisions in association with a new hypothetical gauge boson Z$^{\prime}$. Thus, the search was conducted in the dimuon plus large missing transverse momentum channel. One benchmark scenario of the mono-Z$^{\prime}$, which is known as the Light Vector, has been used for interpreting the CMS open data. No evidence for dark matter has been observed and exclusion limits have been set on the masses of dark matter and Z$^{\prime}$ at 95\% confidence level (CL).

\vspace{0.75cm}
\end{abstract}

\maketitle






\section{Introduction}
\label{sec:intro}
Several cosmological evidences, based on recent observations \cite{planck2015, planck2018, bullet_cluster, R9}, have confirmed the existence of another type of invisible matter which they called the Dark Matter. These studies have also given estimates of the abundance of Dark Matter in the universe to be 27\% of the total energy distribution in the universe.

Several theoretical models were introduced to give a description of the composition of DM. One of the proposed solutions is that dark matter consists of Weakly interacting massive particles (WIMPs). The WIMP theory was successful in giving the correct value of the observed density (WIMP miracle) \cite{R6}. 

Although the standard model of particle physics (SM) provides a decent explanation for the visible matter in the universe, it falls short in providing an explanation for the other type of invisible non-baryonic matter, i.e., dark matter. Other theories beyond the standard model (BSM) are then needed in order to provide a dark matter candidate \cite{SMcource, SMandBSM}.

The masses of the DM candidate can be ranging between few GeV and few 
TeV \cite{Maverick_dark_matter_at_colliders}, which could be produced at the particle colliders, such as the Large Hadron Collider (LHC). 
Many searches for DM have been performed via analysing the data collected by the CMS
experiment during RUN II. These searches rely on the production of a visible object "X", 
which recoils against the large missing transverse momentum from the dark matter particles leaving a signature of $\text{X}+\slashed{p}_{T}$ in the detector . 
The visible particle could be a SM particle like W, Z bosons or jets \cite{R35}, photon \cite{photon} or SM Higgs boson \cite{R36}. 


It is also possible that the visible particle could be a heavy neutral gauge boson (Z$^{\prime}$) predicted by BSM models \cite{R1, R38}.
This type of models is known as Mono-Z$^{\prime}$ model \cite{R1}.
This model contains the following three scenarios, which proposes the production of DM
accompanied by Z$^{\prime}$ boson, which are: Dark Higgs scenario (DH), Light Vector scenario
(LV), and the light Z$^{\prime}$ with inelastic effective field theory (EFT) scenario.
The Z$^{\prime}$ is neutral and can decay leptonically into a pair of oppositely charged leptons (l$^+$l$^-$) or hadronically into a pair of quarks leading to dijet, so that it can be detected as a resonance in the dilepton or dijet invariant mass distribution \cite{R13,R14,R15,R16}. 
The hadronic decay of Z$^{\prime}$ in the Mono-Z$^{\prime}$ portal was studied previously by ATLAS collaboration in \cite{R37}. 

In a previous analysis \cite{EPJplus}, we considered the leptonic decay of Z$^{\prime}$ (i.e. Z$^{\prime}~\rightarrow~\mu^{+}\mu^{-}$), for two scenarios Dark Higgs scenario (DH) and inelastic effective field theory (EFT). The data sets used in this study were obtained from the CMS open data project \cite{R21}, which released data sets from recorded and simulated proton-proton collisions at centre of mass energy ($\sqrt{s} = 8$ TeV). 
These data sets are available publicly for for all researchers even if they are not members in the CMS collaboration. The open data samples provide a great potential for researchers in high energy particle physics to test many theoretical models available in literature \cite{R3}.

The analysis done in the current paper is complementary to our previous work published in \cite{EPJplus}. Here, we present a search for dark fermions (DF), which is originated from Light Vector scenario, in events with dimuon with high invariant mass plus large missing transverse momentum. Similar searches for dark matter in this channel have been performed at the ATLAS and CMS experiments at the LHC with the visible particle being a Z boson decaying to dimuon at $\sqrt{s}$= 8 TeV \cite{R450} and $\sqrt{s}$ = 13 TeV \cite{R45055}.

In the following section \ref{section:model}, the theoretical formalism of the Light Vector scenario and its free parameters are presented. Then the simulation techniques used for events generation for the signal and SM backround samples are displayed in section \ref{section:MCandDat} in addition to the description of CMS open data files from the proton-proton collisions.
The selection cuts and the analysis strategy are then explained in section \ref{section:AnSelection}. 
Finally, the results and the summary of this analysis are discussed in sections \ref{section:Results} and \ref{section:Summary} respectively.

\section{The simplified model}
\label{section:model}
\begin{figure}
    \centering
    
    \includegraphics[width=6.5cm]{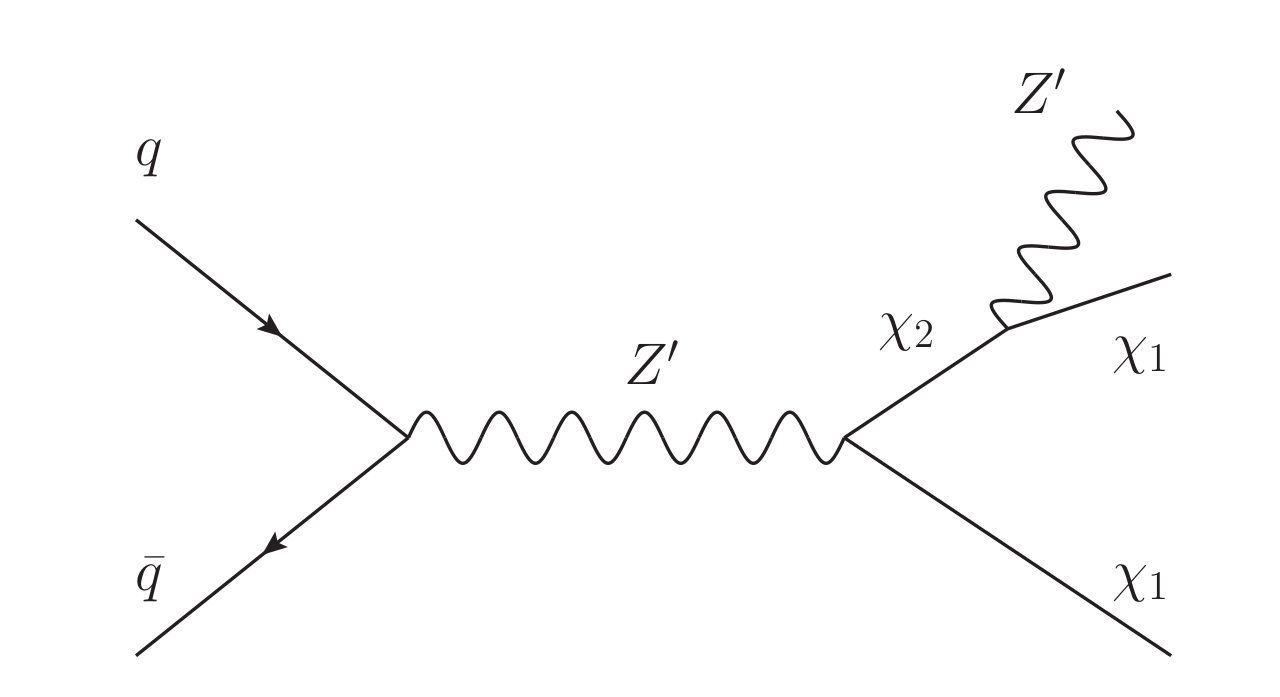}
    \qquad
    \caption{Feynman diagrams for the dark fermion scenario in the mono-Z$^{\prime}$ simplified model.}
    \label{figure:fig1}
\end{figure}

Our target model is known as Mono-Z$^{\prime}$, which has been discussed in \cite{R1}, assumes the production of dark matters from proton-proton collisions at the LHC through a new heavy gauge boson Z$^{\prime}$. 
The dark matter production proceeds through one of three different possible scenarios for the production of dark matter in the Mono-Z$^{\prime}$ model, dark Higgs (DH) scenario and inelastic effective field theory coupling (EFT), which have been previously studied in \cite{EPJplus}.
The third scenario is called Light vector (LV), which is presented in figure \ref{figure:fig1}.

In the rest of this paper we will dedicate our analysis to the LV scenario, which is also know as Dark Fermion (DF) according to \cite{R37}.
The proposed dark fermion can be produced through the process of pair annihilation of two quarks $q\bar{q}$ mediated by the heavy vector boson Z$^{\prime}$, which then undergoes two dark fermions, a light dark fermion $(\chi_{1})$ and a heavy one $(\chi_{2})$. $\chi_{2}$ is heavy enough to decay to a Z$^{\prime}$ and another light dark fermion $\chi_{1}$ (i.e. $\chi_{2}$ $\rightarrow$ Z$^{\prime}$ $\chi_{1}$) as shown in figure \ref{figure:fig1}.

The interaction term, in the Lagrangian, between the dark fermions and Z$^{\prime}$ is given by \cite{R1} 
\begin{equation}
    \frac{\texttt{g}_{DM}}{2} Z^{\prime}_{\mu}\large(\bar{\chi_{2}}\gamma^{\mu}\gamma^{5}\chi_{1} + \bar{\chi_{1}}\gamma^{\mu}\gamma^{5}\chi_{2}\large), \nonumber
\end{equation}
where $\texttt{g}_{DM}$ is the coupling of Z$^{\prime}$ to the dark fermions $\chi_{1}$ and $\chi_{2}$.

Two assumption can be used to set the masses in the dark fermion model. One with a heavy dark sector and the other with a light dark sector as in table \ref{table:tab1}, which are proposed in \cite{R1}.
In the case of the light dark sector case, since the cross section increases with lower $\chi_{1}$ mass, we include optimistic case with very light $\chi_{1} = 1, 5, ..., 50$ GeV, while $\chi_{2}$ is a quite heavier than $\chi_{1}$.
For the case of the heavy dark sector, the dark fermion masses scale with the mediator mass.


\begin{table} [h!]    
\centering
\begin {tabular} {ll}
\hline
\hline
Scenario & \hspace{3pt} Masses assumptions \\
\hline
\\

    & \hspace{4pt} $M_{\chi_{1}} = 1, 5, ...,50$ GeV \\
Light dark sector & \\
    & \hspace{4pt} $M_{\chi_{2}} = M_{\chi_{1}} + M_{Z'} + 25$ GeV
  \\
\hline
\\
& \hspace{4pt} $M_{\chi_{1}} = M_{Z'} / 2$ GeV \\
Heavy dark sector & \\
 &\hspace{4pt} $M_{\chi_{2}} = 2M_{Z'}$ GeV
 \\
\hline
\hline
\end {tabular}
\vspace{3pt}
\caption{The Light and heavy mass assumptions for the dark sector for the dark fermion scenario \cite{R1}.}
\label{table:tab1}
\end{table}
For future reference, the coupling of the Z$^{\prime}$ to the visible fermions will be represented by $\texttt{g}_{SM}$, while the coupling of the Z$^{\prime}$ to the dark fermions will be represented by $\texttt{g}_{DM}$ in the rest of this paper.  The only allowed decays in the DF scenario are assumed to be the decay of  ${Z}'\rightarrow\chi_{1}\chi_{2}$, $\chi_{2}\rightarrow {Z'}\chi_{1}$ and ${Z}'\rightarrow\mu\tilde{\mu}$. Where the total decay width of the ${Z}'$ and ${\chi_{2}}$ can be calculated given the values of the masses of ${Z}'$ and dark fermions and the coupling constants.

The free parameters in this scenario are the lightest dark fermion mass $M_{\chi_{1}}$, the Z$^{\prime}$ mass $M_{Z^{\prime}}$ and the coupling of Z$^{\prime}$ to both SM and DF particles $\texttt{g}_{SM}$ and $\texttt{g}_{DM}$.
The value taken for the coupling constant $\texttt{g}_{SM}$ is 0.1, since the study done in ref.~\cite{R37} has shown that $\texttt{g}_{SM}$ was excluded between 0.13 and 0.26 for dimuon invaraint mass above 200 GeV.
While the value of $\texttt{g}_{DM}$ has been taken to be 1.0, this choice follows the recommendation of the LHC Dark Matter Working Group presented in \cite{DMrecommendationsAtLHC}. 
Meanwhile the values of the masses are not fixed but scanned over.
The signature that these processes leave at the detector, typically consists of two oppositely charged leptons or jets that is produced from the decay of Z$^{\prime}$ in addition to a large missing transverse momentum from the stable dark fermions $\chi_{1}$.
This scenario was previously studied by the ATLAS collaboration in \cite{R37} with the Z$^{\prime}$ decaying hadronically. 
In our study, we have considered the muonic decay of the on-shell Z$^{\prime}$ since the CMS detector is optimized for this decay channel (which is a clean channel with respect to SM backgrounds). 
So that our studied events are with the following topology ($ \mu^{+}\mu^{-} +\slashed{p}_{T}$).
For the dark fermion scenario, with the use of light dark sector case, table \ref{table:tabchi} indicates the cross section times branching ratios calculated for different sets of the Z$^{\prime}$ and $\chi_{1}$ masses. As we can see from the table, the cross section is sensitive to the change in the dark fermion mass. 
The simulated dark fermion signals, used in this analysis, are private production samples, at which we used the matrix element event generator MadGraph5 aMC@NLO v2.6.7 \cite{MG5}. 
We are grateful to Tongyan Lin, co-authors of \cite{R1}, for sharing the so-called Universal FeynRules Output (UFO) for the Mono-Z$^{\prime}$ model.
In the rest of this paper, we will consider the light dark sector scenario and neglect the heavy case, since the cross section times branching ratio calculations, given in table \ref{table:tabchi2}, for heavy dark sector are much lower than the light case by more than factor 10. Hence this analysis does not have any sensitivity to heavy dark sector scenario.
\begin{table*}
\centering
\fontsize{5.pt}{12pt}
\selectfont
\begin{tabular}{|c|c|c|c|c|c|c|c|c|c|c|}
\hline
\Tdiag{.2em}{1.1cm}{$M_{\chi_{1}}$}{$M_{Z'}$}  & 150 & 200 & 250 & 300 & 350 & 400 & 450 & 500 & 600 & 700 \\
\hline
1  & $9.40\times10^{-2}$ & $4.37\times10^{-2}$ & $2.28\times10^{-2}$ & $1.307\times10^{-2}$  & $0.765\times10^{-2}$ & $0.454\times10^{-2}$ & $0.294\times10^{-2}$ & $0.199\times10^{-2}$ & $0.98\times10^{-3}$ & $0.52\times10^{-3}$ \\
\hline
5  & $7.50\times10^{-2}$ & $3.58\times10^{-2}$ & $1.908\times10^{-2}$ & $1.104\times10^{-2}$ & $0.655\times10^{-2}$ & $0.392\times10^{-2}$ & $0.255\times10^{-2}$ & $0.172\times10^{-2}$ & $0.85\times10^{-3}$ & $0.46\times10^{-3}$ \\
\hline
10 & $5.75\times10^{-2}$ & $2.84\times10^{-2}$ & $1.54\times10^{-2}$ & $0.909\times10^{-2}$ & $0.545\times10^{-2}$ & $0.327\times10^{-2}$ & $0.215\times10^{-2}$ & $0.145\times10^{-2}$ & $0.73\times10^{-3}$ & $0.39\times10^{-3}$ \\
\hline
15 & $4.51\times10^{-2}$ & $2.282\times10^{-2}$ & $1.26\times10^{-2}$ & $0.757\times10^{-2}$ & $0.46\times10^{-2}$ & $0.278\times10^{-2}$ & $0.184\times10^{-2}$ & $0.126\times10^{-2}$ & $0.63\times10^{-3}$ & $0.34\times10^{-3}$ \\
\hline
20 & $3.59\times10^{-2}$ & $1.86\times10^{-2}$ & $1.04\times10^{-2}$ & $0.637\times10^{-2}$  & $0.391\times10^{-2}$ & $0.237\times10^{-2}$ & $0.158\times10^{-2}$ & $0.108\times10^{-2}$  & $0.556\times10^{-3}$ & $0.03\times10^{-3}$ \\
\hline
25 & $2.89\times10^{-2}$  & $1.53\times10^{-2}$ & $0.879\times10^{-2}$ & $0.541\times10^{-2}$ & $0.334\times10^{-2}$ & $0.205\times10^{-2}$ & $0.137\times10^{-2}$  & $0.95\times10^{-3}$ & $0.488\times10^{-3}$ & $0.26\times10^{-3}$ \\
\hline
30 & $2.35\times10^{-2}$ & $1.27\times10^{-2}$ & $0.743\times10^{-2}$ & $0.462\times10^{-2}$ & $0.289\times10^{-2}$ & $0.178\times10^{-2}$ & $0.12\times10^{-2}$ & $0.83\times10^{-3}$ & $0.434\times10^{-3}$ & $0.23\times10^{-3}$  \\
\hline
35 & $1.94\times10^{-2}$ & $1.07\times 10^{-2}$ & $0.633\times10^{-2}$ & $0.398\times10^{-2}$ & $0.251\times10^{-2}$ & $0.155\times10^{-2}$ & $0.105\times10^{-2}$ & $0.742\times10^{-3}$ & $0.385\times10^{-3}$ & $0.213\times10^{-3}$ \\
\hline 
40 & $1.61\times10^{-2}$ & $0.909\times10^{-2}$ & $0.543\times10^{-2}$ & $0.343\times10^{-2}$ & $0.218\times10^{-2}$ & $0.137\times10^{-2}$ & $0.936\times10^{-3}$ & $0.657\times10^{-3}$ & $0.343\times10^{-3}$ & $0.192\times10^{-3}$  \\
\hline

50 & $1.14\times10^{-2}$ & $0.66\times10^{-2}$ & $0.407\times10^{-2}$ & $0.26\times10^{-2}$ & $0.16\times10^{-2}$ & $0.106\times10^{-2}$ & $0.739\times10^{-3}$ & $0.371\times10^{-3}$ & $0.278\times10^{-3}$ & $0.157\times10^{-3}$  \\
\hline

\end {tabular}
\caption{ The dark fermion cross section times branching ratios (in pb) calculated for different sets of the masses $M_{\chi_{1}}$ (in GeV), and $M_{Z^{\prime}}$ (in GeV), for the light dark sector mass assumption, with the following couplings constants $\texttt{g}_{SM} = 0.1,~\texttt{g}_{DM} = 1.0$ and at $\sqrt{s} = 8$ TeV.}
\label{table:tabchi}
\end{table*}
\begin{table}
\centering
\selectfont

\begin{tabular}{|c|c|}
\hline
$M_{Z^{\prime}}$ (GeV) & $\sigma \times \text{BR (pb)}$\\
\hline
150  & $1.73\times10^{-2}$ \\
\hline
200  & $0.51\times10^{-2}$ \\
\hline
250 & $0.18\times10^{-2}$ \\ 
\hline
300 & $0.74\times10^{-3}$ \\ 
\hline
350 & $0.32\times10^{-3}$ \\ 
\hline
400 & $0.14\times10^{-3}$ \\
\hline
450 & $0.69\times10^{-4}$ \\ 
\hline
500 & $0.36 \times10^{-4}$ \\ 
\hline
600 & $0.11\times10^{-4}$ \\
\hline 
700 & $0.33 \times10^{-5}$\\ 
\hline
\end {tabular}
\caption{Cross section times branching ratio (in pb) for the heavy dark sector in the DF scenario calculated for different sets of the masses $M_{Z'}$, with the following couplings constants $\texttt{g}_{SM} = 0.1,~\texttt{g}_{DM} = 1.0$ and at $\sqrt{s} = 8$ TeV.}
\label{table:tabchi2}
\end{table}


\section{Data and simulated samples}
\label{section:MCandDat}
\subsection{The CMS detector and reconstructed objects}
The Compact Muon Solenoid is one of four main experiments built to study the proton-proton collision data of the LHC. Located at one of the collision points at the LHC, its main objective is the search for new physics beyond the standard model. The CMS is made of several concentric layers of sub-detectors, each is used for detection of a different kind of particle. The CMS is designed to give a good identification of electrons, photons, hadrons, muons and jets as well as measuring their energy and momentum. 
The technical design of the CMS detector makes it possible to have a good measurement of the missing transverse momentum. A precise measurement of the muon momentum requires a strong magnetic field, hence a super conducting solenoid is used for this purpose. 

The coordinate system of the CMS is designed so that, the origin lies on the collision point. The $x$-axis extends radially from the beamline, the $y$-axis vertically ascends, while the $z$-axis follows the beam's trajectory. The azimuthal angle ($\phi$) describes the particle's angular orientation around the beamline, typically measured in radians. Finally, the pseudorapidity ($\eta$), expressed in terms of the polar angle ($\theta$) is defined as $ \eta = - \text{ln}[\text{tan}(\theta/2)] $.
This way it is possible to calculate the transverse momentum ($p_T$) and transverse energy ($E_T$) from the $x$ and $y$ components of the momentum.

The inner most layer of the detector is the inner tracker, which is used to  measure the momenta of charged particles. The second layer is the Electromagnetic calorimeter (ECAL) 
which is designed for good electron and photon identification and their energy measurements.
The third layer is the Hadron Calorimeter (HCAL) which detects and measures the energy of hardons. The super conducting magnet is fourth layer and it provides a magnetic field of 3.8 T which bends the paths of high energy charged particles allowing to measure their momenta. The outermost layer of the detector is the muon system. The muons system uses three types of detectors: Drift Tubes (DT) in the barrel part of the detector, Cathode Strip Champers (CSC) in the endcaps and Resistive Plate Champers (RPC) completing both the barrel part and endcaps.

Since our study includes muons and missing transverse energy in the final state, we will mention how they are reconstructed. The muon objects are identified and reconstructed from fitting muon tracks from both the inner tracker and the muon system \cite{R18,R40}. 
The missing transverse momentum is reconstructed according to the particle flow (PF) algorithm described in \cite{R19}. The PF algorithm calculates the missing momentum from the imbalance in the vector sum of the momenta in the transverse plane. 
Many factors can affect on the magnitude of the $\vec{\slashed{p}}_{T}$ leading to overestimation or underestimation of its true value. These factors include the calorimeter response, as minimum energy thresholds in the calorimeter and $p_{T}$ thresholds, inefficiencies in the tracker and non-linearity of the response of the calorimeter for hadronic particles \cite{R45}.

To account for the effect of this factors, we replace $\vec{\slashed{p}}_{T}$ by its corrected version $\slashed{\slashed{p}}_{T}^{~\text{corr}}$, which is one of the variables included in the Particle Flow (PF) MET object \cite{R45new, pfmetopendata} in the CMS software \cite{CMSSWversion}.

\subsection{Monte Carlo simulation of the LV scenario}
The signal samples, for LV scenario, have been generated using \text{MadGraph5\_aMC}
\text{@NLO~v2.6.7} \cite{MG5} and the hadronization by Pythia \cite{R34}.  
We produced several samples for the signal at different sets of masses of the dark fermion $\chi_{1}$ and the mediator Z$^{\prime}$ and calculated their cross sections. The range of mass of Z$^{\prime}$ taken was from 150 GeV to 700 GeV. While for the dark fermion Z$^{\prime}$, it was from 1 GeV to 50 GeV. Assuming the couplings are $\texttt{g}_{SM} = 0.1$ and $\texttt{g}_{DM} = 1.0$.

The detector simulation of the read out system response (digitization) and reconstruction processes have been done using the standard CMS open data software framework \cite{CMSSWversion} (the release \text{CMSSW\_5\_3\_32}) at $\sqrt{s} = $ 8 TeV requirements, with the suitable triggers list used for CMS-2012 analysis.
The effect of pile-up has been simulated by overlaying MC generated minimum bias events \cite{pileup}.

\subsection{Backgrounds estimation}
\label{section:background}
The SM background processes yielding lepton pairs in the signal region are the production
of top quark pairs ($\text{t}\bar{\text{t}}$), Drell-Yan (DY) production, and production of diboson (WW, WZ and ZZ).
The second type of background is the jets background, which comes from the misidentification of jets as muons, where a jet or multijet pass the muons selection criteria. 
This kind of backgrounds originate from two processes: W+jet and QCD multijet. 
The contamination of single and multijet background in data is usually estimated from data using a so called data driven method which is explained in \cite{zprime}.  
The third is the cosmic muons background \cite{zprime}.

For simulating background processes, we used the CMS open MC samples at $\sqrt{s}$ = 8 TeV \cite{R39}. 
DY background $q\overline{q} \rightarrow \mu\overline{\mu}$, was generated using \text{POWHEGBox v1.0} MC program \cite{powheg1, powheg2} interfaced to Pythia v.6.4.26 for parton shower model \cite{R34}.
Another important sources of SM backgrounds with dimuon and missing $p_T$ in the final state is the fully leptonic decay of $\text{t}\bar{\text{t}}$, which has been generated using \text{MadGraph5\_aMC@NLO} \cite{MG5}.
The electroweak diboson production channels as WW and WZ were generated with \text{MadGraph} interfaced to
Pythia v.6.4.26, and ZZ to four muons process which is also generated with \text{POWHEGBox v1.0}.
The Monte Carlo samples and their corresponding cross sections used in this analysis, calculated at next-to-leading (NLO) or next-to-next-to-leading order(NNLO), are indicated in table \ref{table:tab3}. 

The contributions of the SM background processes have been estimated from the Monte Carlo simulations, following the same method applied in the previous search for new resonance 
within the dimoun events at $\sqrt{s}= 8$ TeV \cite{zprime}. 
The Monte Carlo sample of the SM backgrounds, which are listed in table \ref{table:tab3}, are normalized to their corresponding cross sections.

It has been noticed that, the contribution of jets background are very small above 400 GeV in the dimuon invariant mass spectrum, as estimated in \cite{zprime}, with only 3 events could be misidentified as muons for an integrated luminosity of 20.6 fb$^{-1}$. 
Thus in our case (luminosity = 11.6 fb$^{-1}$) this contribution is expected to be much lower than 3 events. 
While in a mass bin [120 - 400] GeV, the jets misidentification has been estimated to be 147 events which represents about 0.15\% of the total SM backgrounds (96800 events) estimated in this mass bin \cite{zprime}, which can have a very tiny effect on our results. 
For these reasons QCD and W+jets backgrounds estimated from data are negligible in the current study.

\begin{table*} 
\centering
\begin {tabular} {|l|l|l|c|l|}
\hline
Process \hspace{1cm} & Generator  & Data Set Name & {$\sigma \times \text{BR} ~(\text{pb})$} & Order \\
\hline
\hline
$\text{DY} (\mu\bar{\mu})$  & POWHEG &DYToMuMu\_M-20\_CT10\_TuneZ2star\_v2\_8TeV. \cite{R22}  & 1916 \cite{R3}& NNLO\hspace{6cm}\\
\hline
$\text{t}\bar{\text{t}}$ + jets & MADGRAPH  & TTJets\_FullLeptMGDecays\_8TeV. \cite{R23} & 23.89 \cite{ttbar}& NLO \\
\hline
WW + jets & MADGRAPH & WWJetsTo2L2Nu\_TuneZ2star\_8TeV. \cite{R24} & 5.8 \cite{R3}& NLO \\
\hline
WZ + jets & MADGRAPH  & WZJetsTo3LNu\_8TeV\_TuneZ2Star. \cite{R25} &1.1 \cite{R3}& NNLO \\
\hline
$ZZ\rightarrow 4\mu$  & POWHEG & ZZTo4mu\_8TeV. \cite{R26} & 0.077 \cite{R3}& NLO \\
\hline
\end {tabular}
\vspace{3pt}
\caption{Data sets and their names used for the simulation of the SM backgrounds for proton-proton collisions at $\sqrt{s} = 8$, obtained from the CMS open MC samples, and their corresponding cross sections and order of calculation.}
\label{table:tab3}
\end{table*}

 \begin{table*} 
\centering
\label{ tab-marks }
\begin {tabular} {|c|l|c|}
\hline
Run & Data Set & $\mathcal{L}$ ({fb}$^{-1}$) \\
\hline
\hline
Era B~ & SingleMu/Run2012B-22Jan2013-v1/AOD.\cite{R27}  &  \\ 
   &   & 11.6 \cite{Ropendata}\\
Era C~ & SingleMu/Run2012C-22Jan2013-v1/AOD.\cite{R28}  & \\
\hline
\end {tabular}
\vspace{2pt}
\caption{The CMS-2012 open data samples used in this analysis and the corresponding integrated luminosity.}
\label {table:tab4}
\end{table*}

\section{Selection of event and systematic uncertainties}
\label{section:AnSelection}
In order to select events of interest, a selection criteria was designed to pick events with two high $p_{T}$ muons and large missing transverse momentum resulting from the DF candidate. The selection criteria is as follows:
Each of the two muons should pass the high $(p_{T})$ muon identification introduced in \cite{R41, R32}, and satisfies the following preliminary selection: \\
$\bullet$ $p^{\mu}_{T}$ (GeV) $> 45$, \\ 
$\bullet$ $\eta^{\mu}$ (rad) $<$ 2.1. \\
Thus, the events that were selected are the events with two oppositely charged muons with at least one of them passed the single muon trigger HLT$\_$Mu40$\_$eta2p1. This same selection was applied for the search for new physics in 2012 analysis for events containing dimuon resonance.
Also, we restrict the mass of the dimuon to be over 80 GeV, since the Z$^{\prime}$ mass regime lies above this.

\begin{figure} [h!]
\centering
\resizebox*{9cm}{!}{\includegraphics{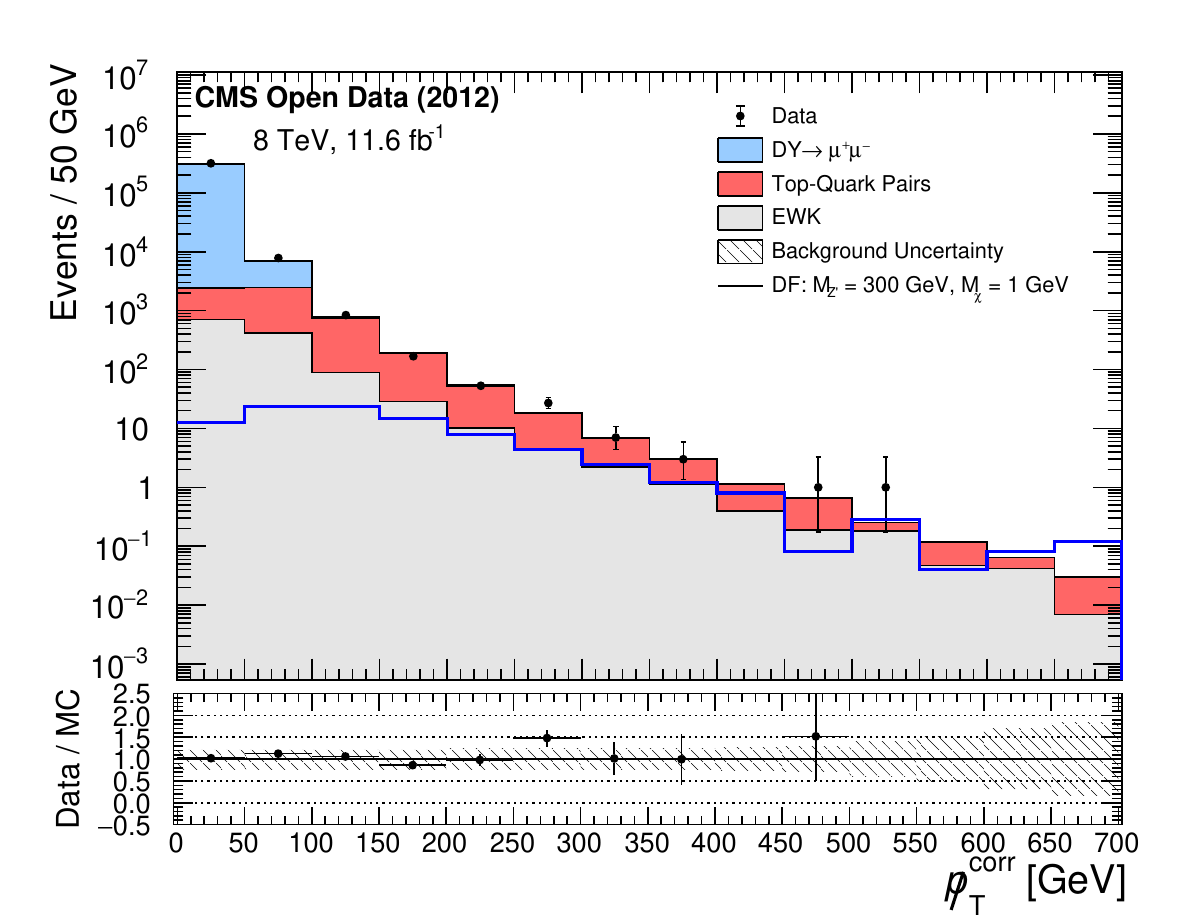}} 
  \caption{The distribution of the missing transverse momentum , after the preliminary selection; for the CMS data, the expected SM backgrounds, and for the DF scenario with $M_{Z^{\prime}} = 300$ GeV and $M_{\chi_{1}} = 1$ GeV.
  In the lower band, the ratio between the data and simulation is shown and 
  the shaded region corresponds to the statistical and systematic uncertainties 
  in the predicted backgrounds, added in quadrature.}
 
\label{figure:fig4}
\end{figure}
Figure \ref{figure:fig4} shows the distribution of the missing transverse momentum, after the preliminary selection; 
the CMS open data are represented by black dots with error bars (statistical error only), the cyan histogram represents the Drell-Yan background, the grey histogram stands for the vector boson pair backgrounds (WW, WZ and ZZ) and the $t\bar{t}$ + jets background is represented by the red histogram. 
These histograms are stacked, while the signal of dark fermion (DF) scenario was generated with invariant mass of the neutral gauge boson Z$^{\prime}$ ($M_{Z^{\prime}}$ = 300 GeV) 
and the mass of the DF ($M_{\chi_{1}}$ = 1 GeV), is represented by the blue colored line, 
and is overlaid.
The total systematic uncertainty (explained in section \ref{section:AnSelection}) is illustrated in the ratio plot in addition to the statistical uncertainties.
This figure shows good agreement between the data points and the simulated SM backgrounds 
within the statistical and systematic uncertainties.

Since the selected signal events, presented by blue solid line in figure \ref{figure:fig4}, are embedded in the background, we need to apply a tighter selection in order to discriminate the signal from the SM backgrounds. This will be explained in the following paragraph.

In addition to the preliminary selection, extra tighter cuts have been applied. 
These tight cuts are based on four variables: the first variable is related to the invariant mass of the dimuon, at which we confined the invariant mass of the dimuon into a small region around the mass of the Z$^{\prime}$, such that $(0.9 \times M_{Z^{\prime}}) < M_{\mu^{+}\mu^{-}} < (M_{Z^{\prime}} + 25)$. 
The second is $\Delta\phi_{\mu^{+}\mu^{-},\vec{\slashed{p}}_{T}^{\text{corr}}}$, which is defined as difference in the azimuth angle between the dimuon direction and the missing transverse momentum direction (i.e. $\Delta\phi_{\mu^{+}\mu^{-},\vec{\slashed{p}}_{T}^{\text{corr}}} = |\phi^{\mu^{+}\mu^{-}}-~\phi^{miss}|$ ), it has been selected to be greater than 2.6 rad.
The Third one is the relative difference between the dimuon$p_{T}$ and the missing transverse momentum ($|p_{T}^{\mu^{+}\mu^{-}} - \slashed{p}_{T}^{\text{corr}}|/p_{T}^{\mu^{+}\mu^{-}}$), it has been selected to be less than 0.6. 
Finally, we apply a requirement on the distance between the two muons in the ($\eta$, $\phi$) plane, $\Delta R < 3$, where $\Delta R = \sqrt{(\Delta\eta)^{2} + (\Delta\phi)^{2}}$.
These tight cuts have been applied in order to strongly decrease the SM backgrounds.

The sources of systematic uncertainties are detector and theoretical related. 
Different sources of the systematic uncertainties, considered in the presented results, are discussed as follows.
The uncertainty in the evaluation of the integrated luminosity of the 2012-data, that are recorded by the CMS, was estimated to be 2.6\% \cite{Lumi}. 
There is 3\% uncertainty related to the detector acceptance and the reconstruction efficiency \cite{zprime}. 
The uncertainty in the energy scale for particles with low energies (unclustered energy) is 10\%, in addition to 2-10\% and 6-15\% uncertainties related to the jet energy scale and jet energy resolution respectively, these uncertainties have a direct impact on the measurement of the missing transverse momentum \cite{R45}.
The uncertainty in the transverse momentum resolution was 5\%, and another 5\% uncertainty from the transverse momentum scale per TeV, due to misalignment in the geometry of the CMS detector \cite{zprime}. 
A 4.5\% uncertainty related to the PDF choice of the DY process \cite{zprime}, was estimated in the present analysis. 5\% uncertainty related to the PDF for the WW process and 6\% for the WZ process, are also included. 
\section{Results}
\label{section:Results}

\begin{figure}[h!]
\centering
  \resizebox*{9cm}{!}{\includegraphics{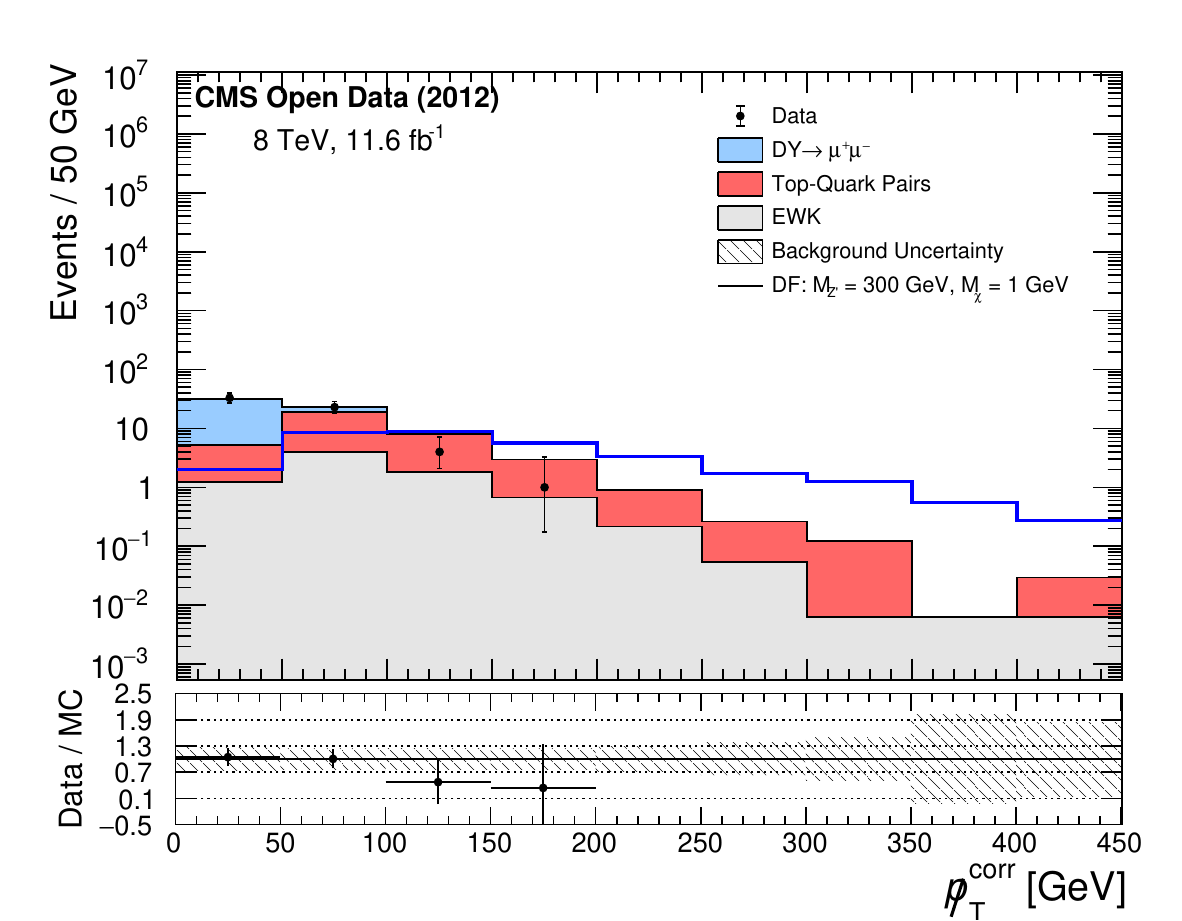}}
  \caption{The distribution of the missing transverse momentum, after final analysis selection cuts, for the expected background and observed events in data in the Z$^{\prime} \rightarrow \mu^{+}\mu^{-}$ channel. One signal benchmark, corresponding to the dark fermion scenario with $M_{Z^{\prime}} = 300$ GeV is superimposed. The signal is normalized to the product of cross section and $\beta$, where $\beta$ represents the Z$^{\prime} \rightarrow \mu^{+}\mu^{-}$ branching fraction. 
  The statistical and systematic uncertainties, added in quadrature, are presented by the hatched band. 
  The ratios of the data and the sum of all the SM backgrounds are shown in the bottom panel.}

  \label{figure:fig6}
\end{figure}

The analysis procedure employed in this study follows a shape-based analysis, where the discriminating variable used to discriminate the signal from the background is the missing transverse momentum distributions ($\slashed{\slashed{p}}_{T}^{~\text{corr}}$). This is justified by the fact that the $\slashed{\slashed{p}}_{T}^{~\text{corr}}$ is much higher for the signal process than for the background process.
The distribution of the missing transverse momentum, after the application of the final event selection, is illustrated in figure \ref{figure:fig6}. The observed data is in good agreement with the simulated backgrounds within the statistical and systematic uncertainties.
The event yields passing the analysis final selection, for each of the SM backgrounds, the DF model (with $M_{Z^{\prime}}$ = 300 GeV, $M_{\chi_{1}}$ = 1 GeV) and the CMS open data; corresponding to an integrated luminosity of 11.6 fb$^{-1}$ are presented in table \ref{table:tab8}. 
\begin{table}[h!]
\centering
\label{ tab-marks }
\begin {tabular} {|l|c|}
\hline
Process & No. of events \\
\hline
\hline
$\text{DY} \rightarrow \mu^{+} \mu^{-}$ & $30.9 \pm  8.3$  \\
\hline
$\text{t}\bar{\text{t}} + \text{jets}$ & $28.3 \pm 6.8$\\
\hline
$\text{WW} + \text{jets}$ & 7.3 $\pm   1.8$ \\
\hline
$\text{WZ} + \text{jets}$ & 0.7 $\pm   0.2$ \\
\hline
$\text{ZZ} \rightarrow 4\mu$  & $0.04 \pm   0.01$   \\
\hline
Sum Bkgs & 67.2 $\pm  16.2$  \\
\hline
\hline
Data & 61 \\
\hline
\hline
DF signal & 36.3 $\pm   8.8$   \\
(at $M_{Z^{\prime}}$ = 300 GeV) &  \\
\hline
\end {tabular}
\caption{The number of events satisfying the criteria of the events selection are illustrated for each SM background, the CMS open data corresponding to a 11.6 fb$^{-1}$ integrated luminosity and the DF scenario signal with coupling constants $g_{DM} = 1.0$,  $g_{SM} = 0.1$ and $M_{\chi_{1}} = 1$ GeV. The total  uncertainty, including the statistical and systematic components, is indicated.}
\label{table:tab8}
\end{table}

In order to make a statistical interpretation for our results, we preformed frequentist (CL) \cite{CLS} with the profile likelihood-ratio test statistic \cite{R2} to derive exclusion limits on the product of signal cross sections and branching ratio Br($Z^{\prime}$ $\rightarrow \mu\mu$) at 95\% confidence level. 

These limits are performed separately for the different signal hypotheses at different masses of $Z^{\prime}$ and $\chi_{1}$. To obtain the $\pm1$ and $\pm2$ sigma bands around the expected limit, pseudo-experiments with the background only hypotheis was used. Where the nuisance parameters where randomely varied within the post fit constraints of the ML fit to the data.

\begin{figure}[h!]
\centering
  \resizebox*{9cm}{!}{\includegraphics{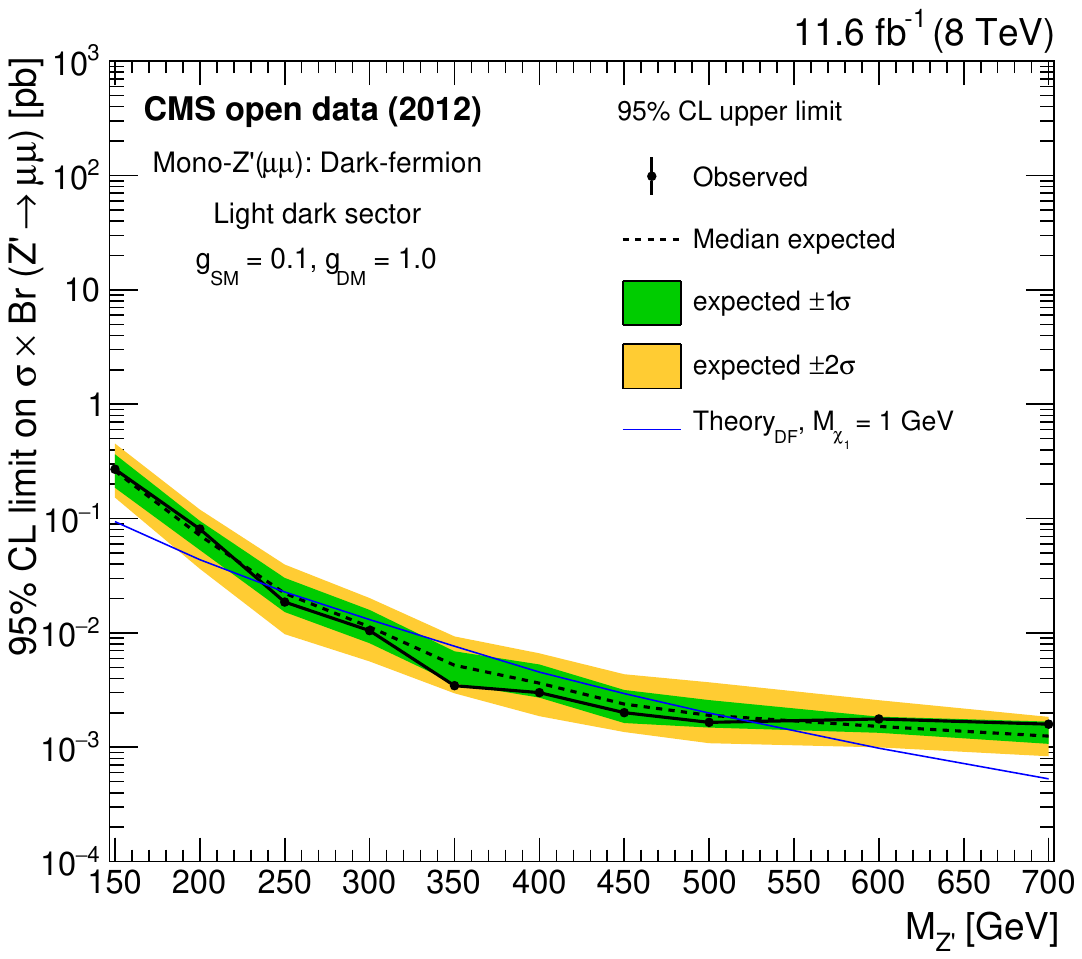}}
  \caption{Observed and expercted upper limits on the cross section times branching ratio as a function of mediator mass for DF mass of $M_{\chi_{1}} = 1$ GeV at 95\% CL. The blue line represents the dark fermion scenario with $M_{\chi_{1}} = 1$ GeV. }
  \label{figure:fig7}
\end{figure}

\begin{figure}[h!]
\centering
  \resizebox*{9.cm}{!}{\includegraphics{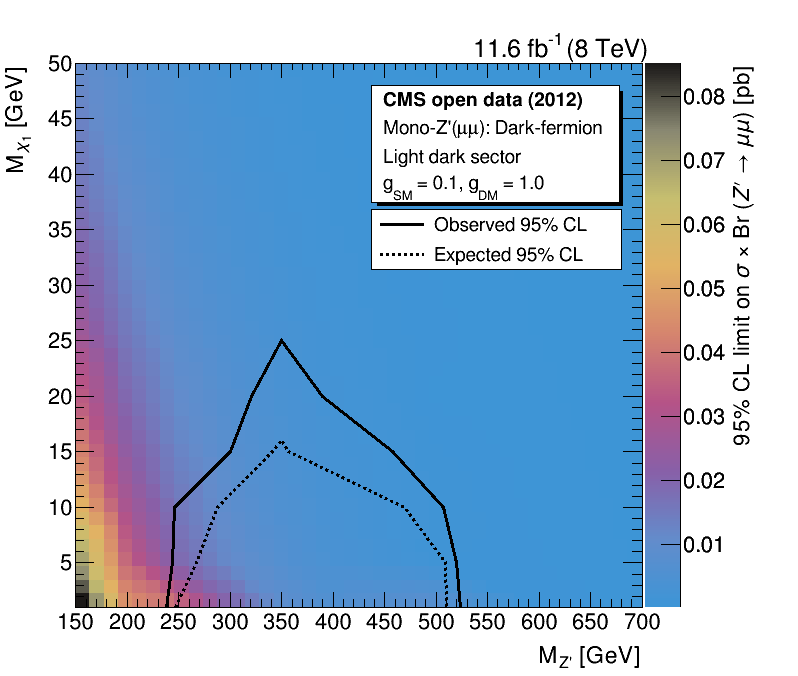}}
   \caption{Exclusion limits at 95\% CL performed on the cross section times branching ratio for variations of pairs of the free model parameters ($M_{Z^{\prime}}$ and $M_{\chi_{1}}$). The filled region indicates the observed upper limit. The solid black curve indicates the observed exclusions for the nominal Z$^{\prime}$ cross section, while the dotted black curve indicates the expected exclusions.}
  \label{figure:fig8}
\end{figure}

 A limit is set on the cross section times the branching ratio Br($Z^{\prime}$ $\rightarrow \mu\mu$) for the dark fermion scenario as shown in figure \ref{figure:fig7}, for the light dark sector masses.
The blue solid line represents the cross section as predicted from theory as a function of $Z^{\prime}$ mass at a fixed dark fermion mass ($M_{\chi_{1}} = 1$ GeV).
No significant deviation from the SM has been observed in any of the studied mass points.
Based on figure \ref{figure:fig7}, $Z^{\prime}$ production is excluded in the mass range between 238 - 524 GeV from the observed data and between 247 - 510 GeV from expected median. 
For the dark fermion scenario, the cross section times the branching ratio limit is presented in figure \ref{figure:fig8}
as a function of the mediator’s masses $M_{Z^{\prime}}$ and the masses of the light dark fermion $M_{\chi_{1}}$. 
The observed exclusion is limited to a narrow region where $M_{\chi_{1}}$ is less than 25 GeV. 

\section{Summary}
\label{section:Summary}
A search for dark fermion particles, based on Mono-Z$^{\prime}$ model, produced in association with a heavy neutral gauge boson Z$^{\prime}$, using set of samples from proton-proton collisions released by CMS open data project corresponding to an integrated luminosity of 
11.6 fb$^{-1}$ during RUN I, has been performed. 
Results from muonic decay mode of Z$^{\prime}$ are discussed, along with its statistical and systematic combination, and presented for the Light Vector (LV) scenario.  
No significant deviation from the standard model prediction has been seen. 
The 95\% CL upper limits on the cross section times the branching ratio (expected and observed), based on the mono-$Z^{\prime}$ model for the LV scenario, were set. 
These limits constitute the most stringent limits on the parameters ($M_{Z^{\prime}}$ and $M_{\chi_{1}}$) of this scenario to date, with fixing the values of the coupling constants to be $g_{DM} = 1.0$,  $g_{SM} = 0.1$.  
For the LV scenario with a light dark sector mass assumption a small region, where the mass of dark fermions ($M_{\chi_{1}}$) is less than 25 GeV, is excluded. 
For $M_{\chi_{1}}$ = 1 GeV, the corresponding excluded range of $M_{Z^{\prime}}$ is 238 - 524 GeV from the observed data and between 247 - 510 GeV from expected median.

\begin{acknowledgments}
Y.Mahmoud wishes to thank the Center for Theoretical Physics (CTP) at the British University in Egypt (BUE) for their continuous support, financially and scientifically, for this work. 
Also we would like to thank Tongyan Lin, co-author of \cite{R1} for providing us with the UFO model files, helping us with generating the signal events and cross-checking the results.
\end{acknowledgments}

\section*{Data Availability Statement}
The data sets used and analysed during the current study are available in the CERN's open data portal.
The URL links:\\
http://opendata.cern.ch/record/7741.\\
http://opendata.cern.ch/record/9577.\\
http://opendata.cern.ch/record/9971.\\
http://opendata.cern.ch/record/9983.\\
http://opendata.cern.ch/record/10071.\\
http://opendata.cern.ch/record/6021.\\
http://opendata.cern.ch/record/6047.\\

\bibliographystyle{spbasic}      
\bibliographystyle{spmpsci}      
\bibliographystyle{spphys}       


\end{document}